# Spherical Coordinates of Carbon Atoms in $C_{20}$ Fullerene Cage

A. S. Baltenkov

Arifov Institute of Ion-Plasma and Laser Technologies,
100125, Tashkent, Uzbekistan

**Abstract**
In the spherical coordinate system with the center in the center of the fullerene cage, the spherical coordinates of 20 carbon atoms forming the $C_{20}$ cage have been calculated.

1. **Introduction**

The consideration of a molecule as a cluster of non-overlapping atomic spheres is a widely used approach in molecular physics despite the fact that this approach is an essential simplification of the real molecular field. This model is usually combined with an additional assumption; beyond the molecular sphere (that encloses the atomic spheres), the molecular field is spherically symmetric and far from the molecule the continuum wave function is the sum of a plane wave plus a single spherical wave emitted from the mass center of the target [1-13]:

$$\psi_{\mathbf{k}}^{+}(\mathbf{r} \to \infty) = e^{i\mathbf{k}\cdot\mathbf{r}} + F(\mathbf{k},\mathbf{k}')\frac{e^{ikr}}{r}. \qquad (1)$$

Here $\mathbf{r}$ is the electron radius vector in the coordinate system with origin at the center of mass, $\mathbf{k}$ and $\mathbf{k}' = k\mathbf{r}/r$ are the initial and final directions of the continuum electron and $F(\mathbf{k},\mathbf{k}')$ is the elastic scattering amplitude:

$$F(\mathbf{k},k') = \frac{1}{2ik}\sum_{l=0}^{\infty}(e^{2i\eta_l}-1)(2l+1)P_l(\cos\vartheta). \qquad (2)$$

The phase shifts $\eta_l(k)$ in Eq.(2) are defined from the matching conditions for the solutions of the wave equation on the surfaces of the atomic and molecular spheres. Thus, the problem of electron scattering by a *spherically non-symmetrical* potential is reduced to the conventional S-matrix method of partial waves for a *spherical scatterer*.

The method developed in papers [14-16] for calculating the continuum wave function differs fundamentally from the above-mentioned traditional consideration of the electron elastic scattering process. According to this approach, in the electron scattering process the atomic spheres become the sources of the scattered waves and far from the cluster of non-overlapping atomic spheres there is a system of spherical waves emerging from each of the spatially separated centers

$$\psi_{\mathbf{k}}^{+}(\mathbf{r}) \approx e^{i\mathbf{k}\cdot\mathbf{r}} + \sum_{j=1}^{N} A_j \frac{e^{ik|\mathbf{r}-\mathbf{R}_j|}}{|\mathbf{r}-\mathbf{R}_j|}. \qquad (3)$$

Here $\mathbf{k}$ is the particle wave vector; the vectors $\mathbf{R}_j$ are the positions of the target centers in an arbitrary coordinate system. Evidently, the difference between the continuum wave functions that have the differing asymptotic forms of equation (1) and equation (3) has consequences. The second case leads to the multi-center picture of scattering, while in the first case, the phenomenon of electron diffraction by molecules as interference of spherical waves becomes impossible because beyond the target, there is only a single spherical wave.

According to [14-16], in the problem of electron scattering by a cluster of atomic potentials, the wave function corresponding to Eq. (3) can be represented as a plane wave and a linear combination of the Green's functions of the free motion and the derivatives of these functions. The boundary conditions imposed on the continuum wave function at the centers of the atomic potentials result in a system of inhomogeneous algebraic equations. Their solution defines the coefficients of the linear combinations and the amplitudes of electron scattering by a target. Thus, in [14-16] it was shown that if one knows the phases shifts for each of the short-range atomic potentials forming the target, and the target geometry, then the amplitude of electron elastic scattering could be written *in closed form* rather than in the form of partial-wave expansion Eq.(2).

## 2. Target geometry

Bearing in mind to further calculate the elastic scattering of slow electrons by fullerenes $C_{20}$, here we will define the target geometry, i.e. in a spherical coordinate system with a center in the center of the fullerene cage we will calculate the spherical coordinates of 20 carbon atoms forming the $C_{20}$ cage.

The $C_{20}$ cage created by the regular pentagons is presented in figure 1. The X-axis of the Cartesian coordinates in this figure is perpendicular to the page plane and crosses the middle of the cage edge 9-12; the Y and Z axes are located in the page plane and cross the middles of 7-15 and 1-2 edges, respectively. The calculation of the spherical coordinates with the polar axis Z can be significantly simplified if one takes into account that the polyhedron vertexes 3, 4, 5, 6 and 13, 14, 19, 17 (black spheres) are the vertexes of cube with the edge length equal to

$$D = r\sqrt{2 - 2\cos 144^o} \approx r \cdot 1.902,$$

where $r$ is the radius of circle in which the regular pentagon is fitted. The center of the cube is coincided with the center of the $C_{20}$ cage; so the fullerene radius $R$ is connected with $r$ as follows:

$$R = D\sqrt{3}/2 \approx r \cdot 1.647.$$

The edge of regular pentagon $L$ is the bound length between carbon atoms and the following relation connects it with $r$

$$L = r\sqrt{2 - 2\cos 72^o} \approx r \cdot 1.176.$$

Eliminating radius $r$ in these formulas, we find that the fullerene radius $R$ is connected with the bound length $L$ by the following way $R \approx L \cdot 1.401$. The bound length $L$ in the $C_{20}$ cage [17] is $L \approx$ 1.456 Å $\approx$ 2.751 $a_0$, where $a_0$ is the Bohr radius; so the $C_{20}$ fullerene radius is $R \approx$ 2.040 Å $\approx$ 3.855 $a_0$ [18]; $D = 4.451$ $a_0$. Spherical angles of carbon atoms in the $C_{20}$ cage are given below in Table 1.

Table 1. Spherical coordinates of carbon atoms in the $C_{20}$ cage; the radial distance $R$ for all atoms is equal to $R \approx$ 2.040 Å $\approx$ 3.855 $a_0$

|    | Polar angle $\theta$, rad | Azimuth angle $\varphi$, rad |
|----|---------------------------|------------------------------|
| 1  | $\theta_1 = \arcsin(L/2R) \approx 0.3648$ | $\varphi_1 = 0$ |
| 2  | $\theta_2 = \theta_1$ | $\varphi_2 = \pi$ |
| 3  | $\theta_3 = \arcsin(2^{-1/2}D/R) \approx 0.9553$ | $\varphi_3 = \pi/4$ |
| 4  | $\theta_4 = \theta_3$ | $\varphi_4 = 3\pi/4$ |
| 5  | $\theta_5 = \theta_3$ | $\varphi_5 = 5\pi/4$ |
| 6  | $\theta_6 = \theta_3$ | $\varphi_6 = 7\pi/4$ |
| 7  | $\theta_7 = \pi/2 - \arcsin(L/2R) \approx 1.2060$ | $\varphi_7 = \pi/2$ |
| 8  | $\theta_8 = \theta_7$ | $\varphi_8 = 3\pi/2$ |
| 9  | $\theta_9 = \pi/2$ | $\varphi_9 = \arcsin(L/2R) \approx 0.3649$ |
| 10 | $\theta_{10} = \pi/2$ | $\varphi_{10} = \pi - \varphi_9$ |
| 11 | $\theta_{11} = \pi/2$ | $\varphi_{11} = \pi + \varphi_9$ |
| 12 | $\theta_{12} = \pi/2$ | $\varphi_{12} = 2\pi - \varphi_9$ |
| 13 | $\theta_{13} = \pi - \theta_3$ | $\varphi_{13} = \pi/4$ |
| 14 | $\theta_{14} = \pi - \theta_3$ | $\varphi_{14} = 3\pi/4$ |
| 15 | $\theta_{15} = \pi/2 + \arcsin(L/2R) \approx 1.9357$ | $\varphi_{15} = \pi/2$ |

| 16 | $\theta_{16}=\pi-\theta_1$ | $\varphi_{16}=0$ |
| 17 | $\theta_{17}=\pi-\theta_3$ | $\varphi_{17}=7\pi/4$ |
| 18 | $\theta_{18}=\theta_{15}$ | $\varphi_{18}=3\pi/2$ |
| 19 | $\theta_{19}=\pi-\theta_3$ | $\varphi_{19}=5\pi/4$ |
| 20 | $\theta_{20}=\pi-\theta_1$ | $\varphi_{20}=\pi$ |

**Acknowledgments**
The author is grateful for the support to the Uzbek Foundation Award ОТ-Ф2-46**.**

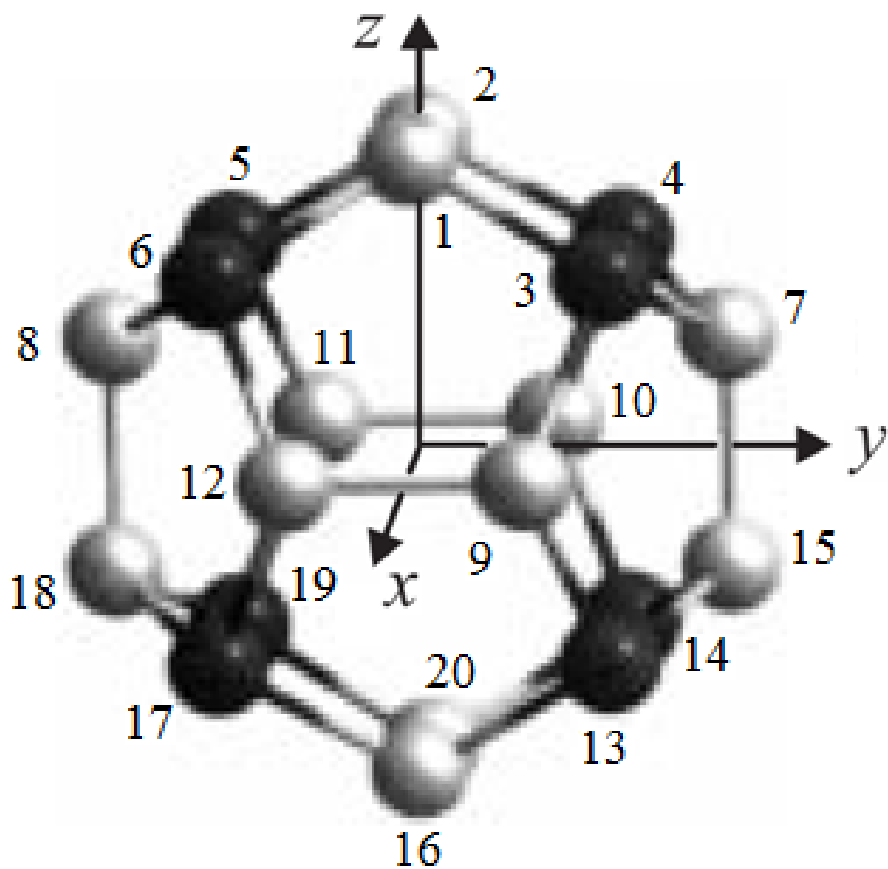

Fig.1. The disposition of 20 carbon atoms in the $C_{20}$ fullerene cage. The Cartesian coordinate systems XYZ; the X-axis is directed perpendicular to the page; Z is the polar axis of the spherical coordinate system. The azimuth angle is measured from the X-axis toward the Y-axis, counterclockwise, viewed from the "North Pole" of the $C_{20}$ sphere.